# TENSOR $^2H^4He$ INTERACTION AND $^6Li$ PROPERTIES IN THE POTENTIAL CLUSTER MODEL

**S.B. Dubovichenko**
Kazak state national university, Almaty, Kazakhstan,

A new parameterization of the gaussian $^2H^4He$ potentials with forbidden states and tensor component, which correctly reproduce scattering phase shift at low energies is offered. On the basis these interactions in a two - cluster model it is possible to describe all main characteristics of a $^6Li$ nucleus bound state, including a sign and value of a quadrupole moment.

In the beginning 70th years in work [1] was shown, that the phase shift of elastic scattering of light cluster systems can be described on the basis deep purely attracting potentials, which contain binding forbidden states (FS). The FS structure is determined by a permutation symmetry of wave functions (WF) system relatively nucleons rearrangements. The behavior of scattering phase shift at zero energy for the given type of interactions submits to the generalized Levinson theorem [2]:

$$\delta_L = \pi ( N_L + M_L )$$

Where $N_L$ and $M_L$ number of forbidden and allowed - really exists states. The phase shift at large energies tend to zero all time remaining positive. Such approach, apparently, it is possible to consider, as alternative to the frequently used concept of the repulsive core, which is introduced for the qualitative take into account the Pauli principle, without complete antisymmetrization WF. The radial WF of allowed states (AS) of potentials with FS oscillates at small distances, but does not die out, as it was for interactions with core.

Then in work [3] were parametrized intercluster gaussian potentials of interaction, correctly reproducing phase shift of $^2H^4He$ elastic scattering at low energies, and containing forbidden states. Is shown, that on the basis such potentials in a cluster model it is possible to reproduce the main characteristics of $^6Li$ nucleus bound states (BS), the clusterization probability which in the considered channel is rather high. Besides all states in such system appear pure on orbital Young schemes [1-3] and potentials, obtained from scattering phase shift, it is possible directly to apply to the description of the ground states (GS) characteristics this nucleus. In a used model it is considered, that the nuclear consists from two unstructured fragments, which can compare properties of appropriate particles in a free state.

Within potential cluster model framework was not taken into account tensor component $^4He^2H$ interactions, which results in appearance D wave in WF BS and scattering and allows to consider a quadrupole moment of the $^6Li$ nuclear. Under tensor potential it is necessary to understand interaction, the operator of which depends on relatively orientation of a total spin of the system and intercluster distance. The mathematical form of such operator completely coincides with an operator of the two-nucleon system, therefore potential by analogy we shall name tensor [4-6].

Apparently, for the first time tensor potentials were used for the description $^{2}H^{4}He$ interaction in the beginning 80-th in work [4], where attempt to enter tensor a component in an optical potential is undertaken. It has allowed appreciably to improve quality of the description of scattering differential cross sections and polarizations. In work [5] on a basis "folding model" accounts of cross sections and polarizations are fulfilled and take into account tensor component of a potential has allowed to improve their description. Hereinafter such approach was used in work [6], where on a basis "folding model" of the NN potentials were obtain $^{2}H^{4}He$ interactions with tensor component. Is shown, that it is basically possible correctly to describe the main characteristics of a $^{6}Li$ bound state, including a correct sign and order of the value of a quadrupole moment.

However, in work [4,5] with obtained potentials only processes of clusters scattering were considered. In [6] the characteristics BS of the $^{6}Li$ nuclear is considered without the analysis of phase shift or cross sections. Nevertheless, Hamiltonian of interaction for $^{2}H^{4}He$ system should be uniform and for processes of scattering and for clusters BS, as it was made in [3] in a case purely central potentials. The large probability $^{2}H^{4}He$ clusterization in $^{6}Li$ nuclear allows to hope for a correctness such task in a potential cluster model.

Correct sign of the quadrupole moment - Q were obtained in work [7], where were used phenomenological wave functions of the nuclear BS. This outcome shows, that and in a simple two-cluster model it is possible, basically, to receive the correct description of a quadrupole moment simultaneously with other characteristics $^{6}Li$. The correct sign of a quadrupole moment with the value -0.076 $Fm^{2}$ were obtained and in some accounts on a method resonance groups [8].

Continuing work in this direction we shall consider influence of effects, which give tensor interaction in potential two-cluster $^{2}H^{4}He$ models of the $^{6}Li$ nuclear with a potential in the form

$$V(r) = V_c(r) + V_t(r) S_{12}, \quad S_{12} = [6(Sn)^2 - 2S^2], \qquad (1)$$
$$V_c(r) = -V_0 \exp(-\alpha r^2), \quad V_t(r) = -V_1 \exp(-\beta r^2).$$

Here S - complete spin of the system, n - single vector, conterminous on a direction to a vector of intercluster distance, $\hbar^2 / m_N = 41.4686$ MeV $Fm^{2}$. Thus, we shall search not parameters of a phenomenological WF BS, as it is made in [7], and intercluster potentials, which would allow correctly to transfer all characteristics of a nuclear bound state and phase shifts of elastic $^{2}H^{4}He$ scattering at low energies.

Thus, shall undertake attempt to describe a continuous and discrete spectrum of the $^2H^4He$ cluster system on the basis a uniform intercluster potential, containing tensor a component.

As GS $^6Li$ [1-3,9] the orbital scheme {42} is compared, in S state should be FS with the scheme {6}. In too time in D wave FS is away, as the scheme {42} is compatible with an orbital moment 2. It signifies, that WF S states will be had node, and WF for D wave - have not node. Such classification of forbidden and allowed states on Young scheme as a whole allows to determine a common forms WF BS.

In accounts we proceed from the system of the usual equations with coulomb term [10]

$$\begin{aligned} & u''(r) + [k^2 - V_c(r) - V_{cul}(r)]u(r) = \sqrt{8}V_t(r)w(r) \quad , \\ & w'' + [k^2 - V_c(r) - V_{cul}(r) - 6/r^2 + 2V_t(r)]w(r) = \sqrt{8}V_t(r)w(r) \end{aligned} \qquad (2)$$

Where $\mu$ - reduced mass, $V_{cul}(r) = 2\mu/\hbar^2 Z_1 Z_2/r$ , $k^2 = 2\mu E/\hbar^2$- wave number of a clusters relative movement. At solution of this system the scattering phase shift $\delta_\alpha$, $\delta_\beta$ and mixing parameter $\varepsilon_1$ are determined.

For bound states of the two-cluster system the following form of boundary conditions is used

$\chi_0 = C_1 u_1 + C_2 u_2 = \exp(-kr)$ ,

$\chi_2 = C_1 w_1 + C_2 w_2 = (1 + 3/kr + 3/(kr)^2)\exp(-kr)$ ,

The replacement of exponential form on exact Whitteker functions resulted in a modification of bound energy not more, than on a few units fifth of a sign after a comma.

In accounts the $^6Li$ quadrupole moment was calculated in view of a deuteron moment $Q_d$ as follows [6]

$$Q = Q_d + Q_0 \quad , \qquad Q_0 = \sqrt{\frac{16\pi}{5}} C_{\alpha d} \sum_{LL'} I_{LL'} R_{LL'} \quad ,$$

where

$$C_{\alpha d} = \frac{Z_\alpha M_d^2 + Z_d M_\alpha^2}{M^2} \quad , \qquad R_{LL'} = \langle \chi_L | r^2 | \chi_{L'} \rangle \quad , \qquad \Phi_L = \chi_L / r \quad ,$$

$$I_{LL'} = (-1)^{J+L'+S}\sqrt{\frac{5(2L+1)(2J+1)}{4\pi}}(L020|L'0)(JJ20|JJ)\begin{Bmatrix} L & S & J \\ J & 2 & L' \end{Bmatrix} .$$

Here $\chi_L$ - radial WF, L and L' - can accept values 0 and 2, Z, M - charges and masses of clusters and nuclear. As a deuteron quadrupole moment is 0.286 $Fm^2$ [6], for deriving a correct moment $^6Li$, which is equal -0.0644 (7) $Fm^2$ [11] it is necessary that the value $Q_0$ was -0.3504 $Fm^2$. In the final for $^6Li$ we obtain

$$Q_0 = \frac{4\sqrt{2}}{15}\int r^2(\chi_0\chi_2 - \frac{1}{\sqrt{8}}\chi_2^2)dr$$

The quadrupole moment can be determined and on a basis coulomb form factor as

$$Q_f = 9\sqrt{2}\lim_{q\to 0}\left(\frac{F_{C2}(q)}{q^2}\right)$$

The impulse distribution of clusters, normalized per unit at momentum transferred q = 0, was defined as [2]

$$P^2 = \sum_L P_L^2(q) \quad , \qquad P_L(q) = \int \chi_L j_L(qr)dr \quad .$$

Here q - momentum transferred, $j_L$ - Bessel function, L = 0,2.

Magnetic moment of the nuclear in the two-cluster system in a case, when only one from clusters has a magnetic moment $\mu_d$ and the spin 1 can be represented [9]

$$\mu = \mu_d J + \frac{1}{2(j+1)}(B_{\alpha d} - \mu_d)(\hat{J} + \hat{L} - \hat{S})P_D \quad , \qquad \hat{A} = A(A+1) \quad ,$$

Where $P_D$ - value of impurity D state, L=2 and

$$B_{\alpha d} = \frac{1}{M}\left(\frac{Z_\alpha M_d}{M_\alpha} + \frac{Z_d M_\alpha}{M_d}\right) .$$

The magnetic moment of a deuteron is equal $0.857\mu_0$, and for $^6Li$ nuclear is 0.822 $\mu_0$. Therefore, for deriving in a considered model of a correct moment of the nuclear it is necessary to accept about 6.5 % of impurity D state.

At accounts coulomb form factors was used expression [3,9]

$$F^2 = \frac{1}{Z^2}\sum_J V_J^2 \quad , \qquad V_J = Z_1 F_1 I_{2,J} + Z_2 F_2 I_{1,J} \quad ,$$

Where the integrals from radial functions BS are represented as

$$I_{k,0} = \int (\chi_0^2 + \chi_2^2) j_0(g_k r) dr \quad , \qquad I_{k,2} = 2\int \chi_2 (\chi_0 - \frac{1}{\sqrt{8}}\chi_2) j_2(g_k r) dr \quad .$$

Here k=1 or 2 designates $^2H$ or $^4He$, $g_k= (M_k/M)q$, J - multipolarity of a form factor, equal 0 or 2. The form factors of clusters $F_1$ and $F_2$ are represented in the following parametrization [3,9]

$F_\alpha=(1 - (aq^2)^n)\exp(-bq^2)$ ,

Where a=0.09985 $Fm^2$, b=0.46376 $Fm^2$ and n=6. For deuteron other form was used

$F_d=\exp(-aq^2)+bq^2\exp(-cq^2)$ ,

With parameters a=0.49029 $Fm^2$, b=0.01615 $Fm^2$ and c=0.16075 $Fm^2$.

For calculation of asymptotic constants known expressions [3,9,12] were used

$$\Phi_L = \frac{\sqrt{2k}}{r} C_L^0 A_L \exp(-kr) \quad , \quad A_0 = 1 \quad , \quad A_2 = [1 + 3/kr + 3/(kr)^2] \quad ,$$
$$\Phi_L = \frac{\sqrt{2k}}{r} C_L^{w0} A_L W^0_{-\gamma,l+1/2}(2kr)$$
$$\Phi_L = \frac{\sqrt{2k}}{r} C_L^{w} W_{-\gamma,l+1/2}(2kr)$$

Where k - wave number, defined of the nuclear bound energy in the considered channel, $\gamma$ - coulomb parameter,

$$W_{-\gamma,l+1/2}(2kr)$$

is Whitteker function and

$$W^0_{-\gamma,l+1/2}(2kr) = (2kr)^{-\gamma}\exp(-kr)$$

- its asymptotic. For relation of the asymptotic constants used usual definition

$$\eta_D = C_2^0 / C_0^0 \quad , \qquad \eta_{D1} = C_2^{w0} / C_0^{w0} \quad , \qquad \eta_{D2} = C_2^{w} / C_0^{w} \quad .$$

At such representation at once it is visible, that $\eta_D = \eta_{D1}$.

Radius of the nucleus was calculated by two ways. In a cluster model, where it is possible to present as [3,9]

$$R_r^2 = \frac{M_d}{M} R_d^2 + \frac{M_\alpha}{M} R_\alpha^2 + \frac{M_d M_\alpha}{M^2} R_{\alpha d}^2 \quad , \qquad R_{\alpha d}^2 = \int r^2 (\chi_0^2 + \chi_2^2) dr \quad .$$

And through coulomb form factor [9]

$$R_f^2 = 6 \lim_{q \to 0} \left( \frac{1 - F_{C0}(q)}{q^2} \right) \quad .$$

Here for radiuses of clusters values $R_\alpha$=1.67 Fm, $R_d$=1.96 Fm were used.

In known three-body accounts [13,14] almost all considered characteristics of the $^6$Li nucleus are reproduced practically correctly. However the quadrupole moment it appears positive, also, as well as $\eta_D$= 0.018 - 0.055 is positive, but the experimental value 0.005 $\pm$ 0.017 is resulted in [12] accepts negative values. The asymptotic constant $C_0^0$ in various variants three-body accounts [14] is in the field of 2.2 - 2.4 at experimental value 2.15(10) [12]. In work [13] for $C_0^W$ value 2.71 is obtained at a normalization WF on 70-75 % probability in $^2$H$^4$He the channel. In [14] for probability $^2$H$^4$He channel in S state is found 60-65 %, and the probability for D state in this channel appear very small 0.025-0.63 %. Charging radius in three-body accounts is in limits 2.26-2.43 Fm and a little bit less than experimental values 2.56 (5) and 2.54 (6) Fm [8].

For deriving a negative sign of a quadrupole moment it is necessary, apparently, and negative value $\eta_D$. The possibility negative $\eta_D$ is marked in work [15], where the new experimental data are resulted which give the value - (0.01 $\pm$ 0.015), that it is possible only at different signs in an asymptotic S and D parts of WF. Just such aspect WF BS was obtained in work [6], that has allowed to receive a correct negative sign of a quadrupole moment.

From whole told becomes above clearly, which common form should have WF, correctly to describe a quadrupole moment. Most likely, tensor the part of a potential should be rather narrow that there was no node in D wave. And the central part will be wide with depth, capable to ensure presence of a node in S wave. As a whole tensor the interaction will rather weak and consequently at construction of a potential it be possible emanate from outcomes of work [3], where parameters of the central gaussian potential $V_0$=76.12 MeV and $\alpha$=0.2 Fm$^{-2}$ are resulted.

On the basis these representations two variants of a potential parameters were obtained and are resulted in Tab. 1. The interactions parameters were coordinated to the BS energy, nuclear quadrupole moment and scattering phase shifts.

Table 1. Parameters of the tensor potentials of $^2H^4He$ interaction.

| # | $V_0$ , MeV | $\alpha$ , $Fm^{-2}$ | $V_1$ , MeV | $\beta$ , $Fm^{-2}$ |
|---|---|---|---|---|
| 1 | 71.979 | 0.2 | 27.0 | 1.12 |
| 2 | 77.106 | 0.22 | 40.0 | 1.6 |

The calculation results of the characteristics $^6Li$ GS for these interactions are resulted in Tab. 2 together with the experimental data from [11,12,15]. It is visible, that obtained potentials allow in basic correctly to describe considered experimental results, and the parameters of a central part little differ from parameters of purely central interaction [3]. We shall mark, that if to consider only characteristics BS, it is possible to find very many variants of parameters, which allow to describe all characteristics. And only considering simultaneously the scattering phase shift and BS properties it is possible practically simple to fixed potential parameters.

Table 2. Results of calculations in two-cluster $^2H^4He$ model and experimental [11,12,15] values for the characteristics of $^6Li$ bound state.

| Nuclear characteristics | Accounts for potential #1 | Accounts for potential #2 | Experimental data |
|---|---|---|---|
| $E_{bs}$ , MeV | -1.4735 | -1.4735 | -1.4735 |
| $R_r$ , Fm | 2.599 | 2.563 | 2.56(5); |
| $R_f$ , Fm | 2.532 | 2.496 | 2.54(6) |
| Q , $Fm^2$ | -0.064 | -0.064 | -0.0644(7) |
| $Q_f$ , $Fm^2$ | -0.064 | -0.064 | |
| $C_0^0$ | 1.9(1) | 1.9(1) | 2.15(10) |
| $\eta_D$ | -0.0115(5) | -0.0120(5) | -0.0125(25) |
| $\eta_{D2}$ | -0.0119(3) | -0.0122(2) | 0.005(17) |
| $C_0^{wo}$ | 2.97(3) | 2.85(5) | --- |
| $C_0^{w}$ | 3.17(3) | 3.03(3) | --- |
| $\mu_d/\mu_0$ | 0.848 | 0.847 | 0.822 |
| $P_D$ , % | 1.59 | 1.78 | --- |

The charging radiuses $R_r$ and $R_f$, calculated on the basis these potentials, differing on 2-3 % are in an interval of experimental errors. The asymptotic constant $C_0^W$ = 3.0 -3.2 will be well coordinated with calculations in a microscopic model [16], where the value 3.3 is obtained and with three-body accounts, if to enumerate resulted in [13] a constant on single probability $^2H^4He$ of the channel - 3.1 - 3.2. For probability D state the value a little bit smaller is found, than it is required for the correct description of a magnetic moment. However, the similar situation exists and in classical NN task. For deriving a deuteron magnetic moment the impurity D state about 4 % is required, and phenomenological potentials give the value about 6 -7 % [17].

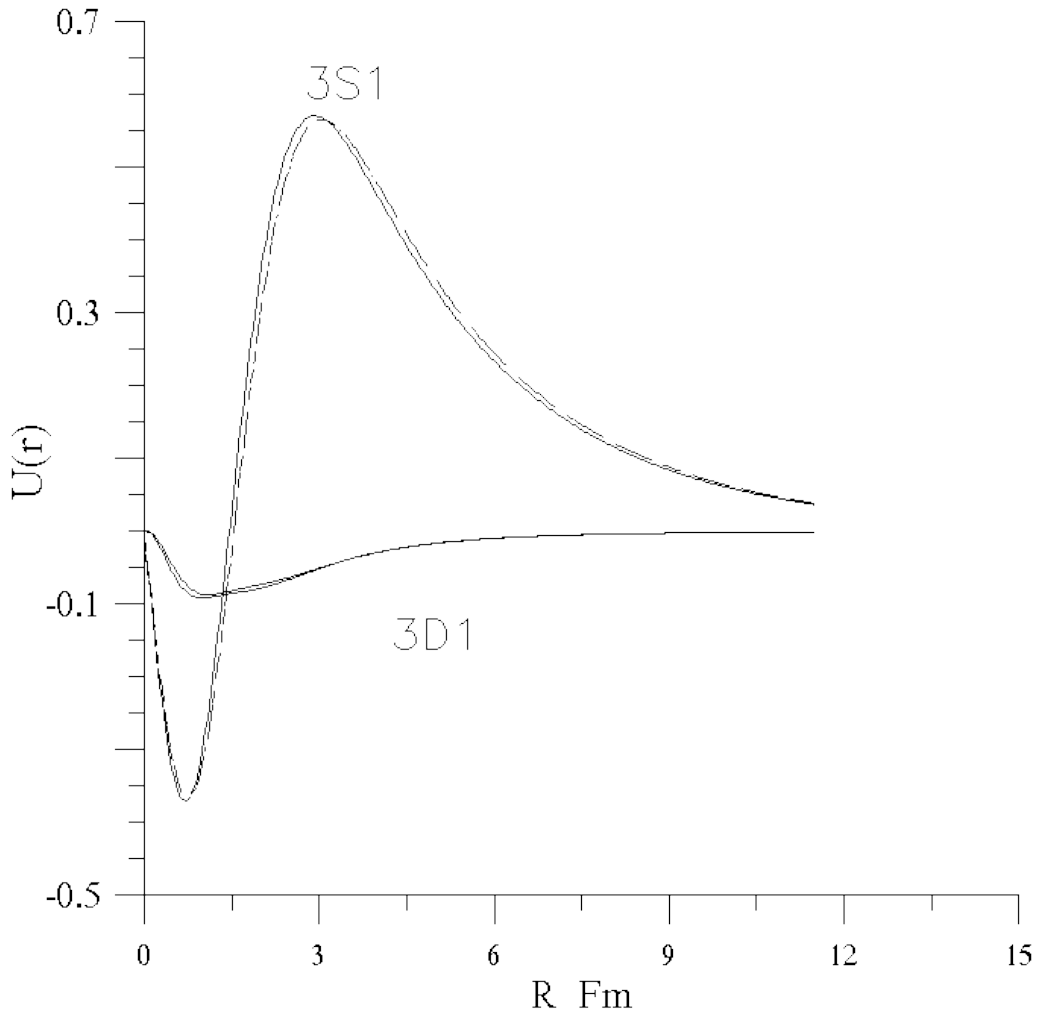

Fig.1. BS WF.

On fig.1 the WF of the BS for the received potentials are shown. (A continuous line for first and dashed for the second variants.). On fig. 2 by a continuous line is shown a coulomb form factor $^6Li$ for the first variant of a potential and dashed for second. The dot-dashed line is results the contribution C2 component to a complete form factor. The experimental data are taken from work [18].

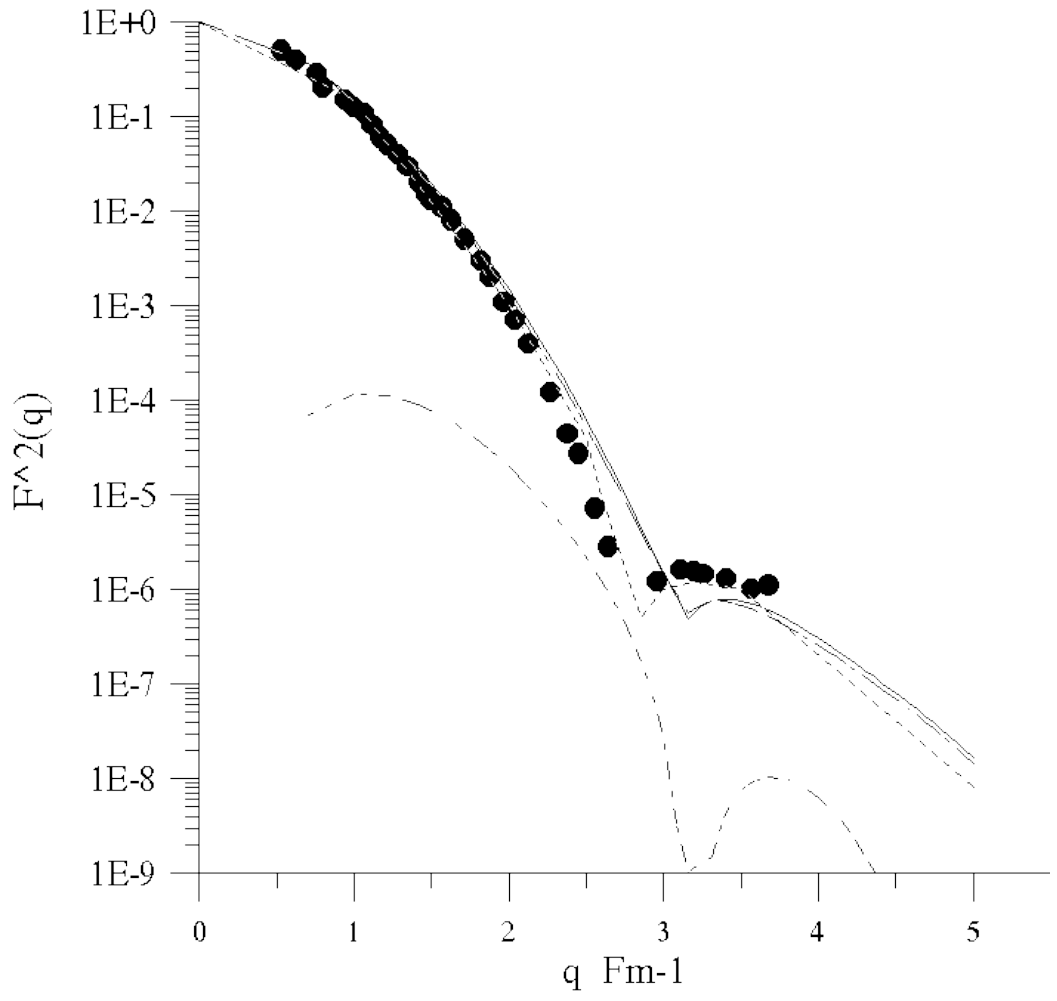

Fig.2. Coulomb form factor $^6Li$.

It is visible, that the contribution C2 in a form factor is rather small and essentially does not change outcomes of accounts in comparison with purely central interaction, which gives only C0 item. The short dashed line on fig. 2 shows outcomes of account for a form factor in three-body model with a complete antisymmetrization WF [19], which practically correctly describe the experimental data.

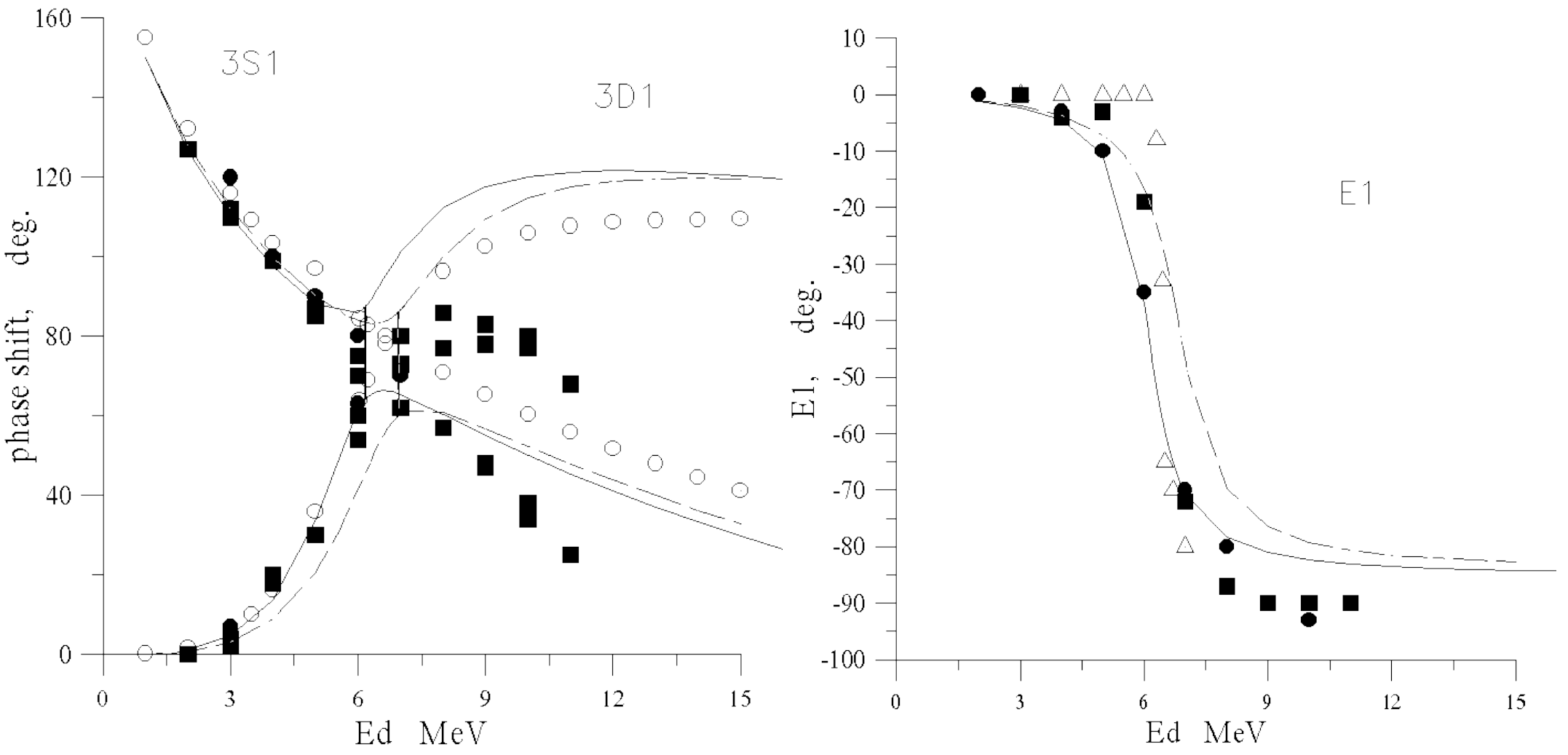

Fig.3. Phase shift and mixing parameter of elastic scattering $^2H^4He$.

On fig. 3 phase shift of elastic scattering and mixing parameter in comparison with the experimental data [21] are shown. We shall mark, that there are and other data [22], there the scattering parametrization are different from representation Blatt-Biedanharn. From fig.3 it is visible, that at small energies it is quite possible to transfer a behavior of experimental scattering phase shift. The negative value of the mixing parameter can be received only at narrow tensor a potential. Wider potentials result in change of a sign $\varepsilon_1$ and simultaneously with it changes a sign S WF BS.

On fig. 4 the solids line is results the impulse distributions $^2H^4He$ clusters in the $^6Li$ nuclear for first and dashed for the second variants of interaction, together with the experimental data [20]. The dot line shows the contribution $P_2$ item, which smoothes a minimum in impulse distribution in region 0.7-0.8 $Fm^{-1}$, not resulting to essential differences from outcomes for central interaction.

From obtained results it is visible, that within the framework of a potential cluster model it is possible to receive potentials $^2H^4He$ interaction with tensor component, permitting correctly to transfer all considered characteristics of a bound state $^6Li$ and phase shift of elastic $^2H^4He$ scattering at low energies. The a little value of a form factor at large momentum transfer can be, apparently, is explained by absence in the present accounts take into account of exchange effects, as it is made in [19] on a basis three-body model.

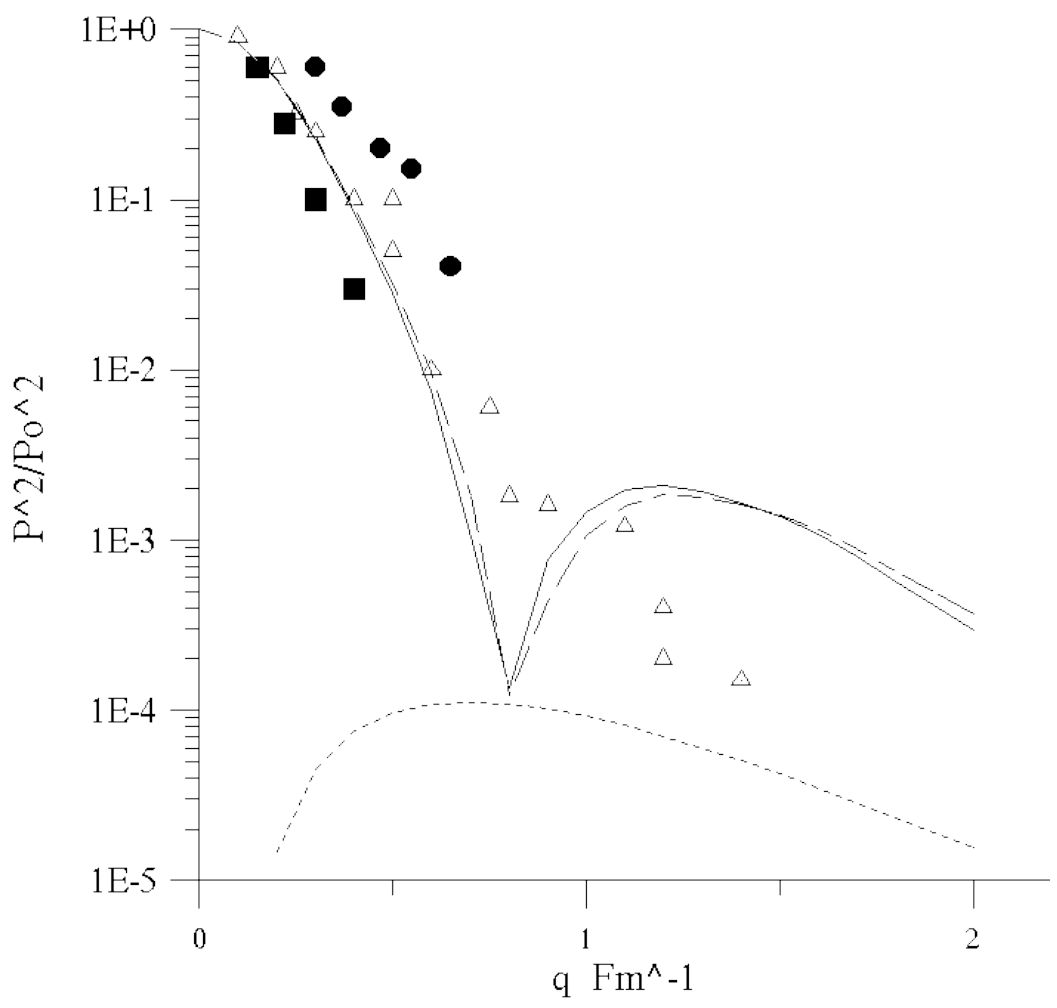

Fig.4. Impulse distributions $^2H^4He$ clusters in the $^6Li$.